\documentclass[12pt]{article}
\usepackage{float} 
\usepackage{amsmath}
\usepackage{slashed}
\usepackage{amsfonts}   
\usepackage{amssymb}

\begin{document}
\date{}
\title{{\bf{\Large Nonrelativistic strings on $ R \times S^2 $ and integrable systems}}}
\author{
 {\bf {\normalsize Dibakar Roychowdhury}$
$\thanks{E-mail:  dibakarphys@gmail.com, dibakar.roychowdhury@ph.iitr.ac.in}}\\
 {\normalsize  Department of Physics, Indian Institute of Technology Roorkee,}\\
  {\normalsize Roorkee 247667, Uttarakhand, India}
\\[0.3cm]
}

\maketitle
\begin{abstract}
We show that the (torsional) nonrelativistic string sigma models on $ R\times S^2 $ can be mapped into \emph{deformed} Rosochatius like integrable models in one dimension. We also explore the associated Hamiltonian constrained structure by introducing appropriate Dirac brackets. These results show some solid evidence of the underlying integrable structure in the nonrelativistic sector of the gauge/string duality.  
\end{abstract}
Over the past one decade, a considerable amount of attention has been paid towards the formulation of nonrelativistic (NR) string sigma models \cite{Gomis:2000bd}-\cite{Gomis:2005pg} on curved manifolds with \emph{local} Galilean invariance. Typically, these sigma models could be classified into two different categories - (1) sigma models over String Newton-Cartan (SNC) geometry \cite{Bergshoeff:2019pij}-\cite{Gomis:2019zyu} and, (2) sigma models over torsional Newton-Cartan (TNC) geometry \cite{Harmark:2017rpg}-\cite{Roychowdhury:2020kma}.

Studying string theory in either of these geometries could in principle lead to several interesting consequences among which two are centrally important - (1) the emergence of a UV complete theory of nonrelativistic quantum gravity (that has found directions until very recently in the context of bosonic sigma models \cite{Gomis:2019zyu},\cite{Gallegos:2019icg}) and, (2) deeper understanding of the underlying \emph{integrable} structure (if any) in the context of nonrelativistic sigma models. This paper aims to gain further insights along the second line of thought. 

The question that we would like to pose in this paper is whether NR strings are \emph{integrable} or not. This question has standard/traditional answers \cite{Arutyunov:2003uj}-\cite{Hernandez:2015nba} in the context of $ AdS_5 \times S^5 $ (super)string theory is by reducing the sigma model into 1D and thereby mapping it to Neumann-Rosochatius integrable models \cite{neumann1}-\cite{Babelon:1992rb}. Here, we are interested in exploring whether similar reduction is possible for NR sigma models on $ R \times S^2 $.

We address these issues explicitly by constructing NR sigma models corresponding to spinning string configurations over $ R \times S^2 $.  We start by writing down the generic $ (d+2) $ dimentional Lorentzian metric\footnote{Here, $ X^{\mathfrak{u}}=\mathfrak{u} $ is the null isometry direction associated with the 4D Lorentzian spacetime.} \cite{Harmark:2017rpg},
\begin{eqnarray}
ds^2 = 2\tau (d\mathfrak{u}-\mathfrak{m})+\mathfrak{h}_{\mu \nu}dX^{\mu}dX^{\nu}
\label{e1}
\end{eqnarray}
where we identify each of the individual one forms\footnote{Here, $ v=t+\frac{\psi}{2} $ is the time coordinate for the reduced TNC spacetime \cite{Grosvenor:2017dfs}.} as \cite{Grosvenor:2017dfs},
\begin{eqnarray}
\tau &=& dt + \frac{1}{2}d \psi - \frac{1}{2}\cos\theta d\varphi = dv - \frac{1}{2}\cos\theta d\varphi \\
\mathfrak{m}&=&\frac{1}{4}\cos\theta d\varphi
\end{eqnarray}
together with the metric on the two sphere,
\begin{eqnarray}
ds^2_{S^2}=\mathfrak{h}_{\mu \nu}dX^{\mu}dX^{\nu}&=& d\theta^2 +\sin^2\theta d\varphi^{2}.
\end{eqnarray}

As a next step, we set $ X^0=t=\kappa \tau $ and introduce the \emph{embedding} coordinates,
\begin{eqnarray}
\label{e5}
\mathcal{Z}_1 &=& X^1 + i X^2 = \sin\theta e^{i \varphi (\tau , \sigma )}= \vartheta_{1} (\sigma)e^{i \varphi (\tau , \sigma )}\\
\mathcal{Z}_2 &=& X^3 = \cos\theta = \vartheta_{2} (\sigma)
\label{e6}
\end{eqnarray}
such that, $ ds^2_{S^2}= d\mathcal{Z}_{i}d\bar{\mathcal{Z}}_{i} ~(i=1,2)$ together with the constraint, $\vartheta^{2}_{i}= \vartheta_{1}^{2}+\vartheta_{2}^{2}=1  $. In terms of embedding coordinates ($ X^{m} $ ($ m=1,2,3 $) ) the metric on two sphere has an equivalent representation, $ ds^2_{S^2}= \mathfrak{h}_{mn}dX^{m}dX^{n}$ where $ \mathfrak{h}_{mn}=diag (1,1,1) $. 

Closed strings propagating over (\ref{e1}) are described by the 2D sigma model action of the following form \cite{Harmark:2019upf},
\begin{eqnarray}
\mathcal{S}=\frac{\sqrt{\lambda}}{4 \pi}\int d^{2}\sigma \mathcal{L}
\label{e7}
\end{eqnarray}
where the corresponding Lagrangian density could be formally expressed as,
\begin{eqnarray}
\mathcal{L}=\sqrt{-\gamma}\gamma^{\alpha \beta}\mathfrak{h}_{\mu \nu}\partial_{\alpha}X^{\mu}\partial_{\beta}X^{\nu}-\sqrt{-\gamma}\gamma^{\alpha \beta}(\tau_{\alpha}\mathfrak{m}_{\beta}+\tau_{\beta}\mathfrak{m}_{\alpha})\nonumber\\
+2(\sqrt{-\gamma}\gamma^{\alpha \beta}\tau_{\beta}+\varepsilon^{\alpha \beta}\partial_{\beta}\zeta)\mathcal{A}_{\alpha}.
\label{e8}
\end{eqnarray}
Here, we identify $ \gamma^{\alpha \beta}=\eta^{ab}\mathfrak{e}^{\alpha}~_{a} \mathfrak{e}^{\beta}~_{b}~(a=0,1)$ as 2D Lorentzian metric on the world-sheet together with the determinant $ \sqrt{-\gamma}=| \mathfrak{e}| $. Moreover, the world-sheet scalar $ \zeta (\sigma^{\alpha})$ plays the role of an additional \emph{compact} dimension associated with the null reduced target space geometry. A non zero winding of string along $ \zeta $ is what guarantees the conserved momentum along the null isometry direction of the target space geometry \cite{Harmark:2017rpg},\cite{Harmark:2019upf}.

We expand the abelian one form as \cite{Harmark:2018cdl}, 
\begin{eqnarray}
\mathcal{A}_{\alpha}=\mathfrak{m}_{\alpha}+\frac{1}{2}(\lambda_{+}-\lambda_{-})\mathfrak{e}_{\alpha}~^{0}+\frac{1}{2}(\lambda_{+}+\lambda_{-})\mathfrak{e}_{\alpha}~^{1}
\label{e9}
\end{eqnarray}
where we identify $ \lambda_{\pm} $ as Lagrange multiplier.

Substituting (\ref{e9}) into (\ref{e8}) we find,
\begin{eqnarray}
\mathcal{L}=\sqrt{-\gamma}\gamma^{\alpha \beta}\mathfrak{h}_{\mu \nu}\partial_{\alpha}X^{\mu}\partial_{\beta}X^{\nu}+2\varepsilon^{\alpha \beta}\mathfrak{m}_{\alpha}\partial_{\beta}\zeta\nonumber\\
+\lambda_{+}\varepsilon^{\alpha \beta}\mathfrak{e}_{\alpha}~^{+}(\tau_{\beta}+\partial_{\beta}\zeta)+\lambda_{-}\varepsilon^{\alpha \beta}\mathfrak{e}_{\alpha}~^{-}(\tau_{\beta}-\partial_{\beta}\zeta)
\label{e10}
\end{eqnarray}
where we identify, $ \mathfrak{e}_{\alpha}~^{\pm}=\mathfrak{e}_{\alpha}~^{0}\pm \mathfrak{e}_{\alpha}~^{1} $. The equations of motion corresponding to Lagrange multipliers impose the following set of constraint equations,
\begin{eqnarray}
\varepsilon^{\alpha \beta}\mathfrak{e}_{\alpha}~^{\pm}(\tau_{\beta}\pm \partial_{\beta}\zeta)=0
\label{e11}
\end{eqnarray}
which has a natural solution of the form,
\begin{eqnarray}
\mathfrak{e}_{\alpha}~^{\pm}=(\tau_{\alpha} \pm \partial_{\alpha}\zeta).
\label{E12}
\end{eqnarray}

Substituting (\ref{E12}) into (\ref{e10}) we arrive at the following Nambo-Goto (NG) Lagrangian,
\begin{eqnarray}
\mathcal{L}_{NG}=\frac{\varepsilon^{\alpha \tilde{\alpha}}\varepsilon^{\beta \tilde{\beta}}}{\varepsilon^{\alpha \beta}\mathfrak{e}_{\alpha}~^{0}\mathfrak{e}_{\beta}~^{1}}(\mathfrak{e}_{\tilde{\alpha}}~^{0}\mathfrak{e}_{\tilde{\beta}}~^{0}-\mathfrak{e}_{\tilde{\alpha}}~^{1}\mathfrak{e}_{\tilde{\beta}}~^{1})\mathfrak{h}_{mn}\partial_{\alpha}X^{m}\partial_{\beta}X^{n}+2\varepsilon^{\alpha \beta}\mathfrak{m}_{\alpha}\partial_{\beta}\zeta
\label{e12}
\end{eqnarray}
where for the present configuration we note down,
\begin{eqnarray}
\mathfrak{e}_{\alpha}~^{0} &=&\tau_{\alpha}=\partial_{\alpha}t + \frac{1}{2}\partial_{\alpha}\psi -\frac{1}{2}\cos\theta \partial_{\alpha}\varphi\\
\mathfrak{e}_{\alpha}~^{1} &=&\partial_{\alpha}\zeta
\end{eqnarray}
together with the 2D Levi-Civita convention, $ \varepsilon^{01}=-\varepsilon_{01}=+1 $ both for the world-sheet as well as tangent space indices. Here, $ X^{m} $s ($ m=1,2,3 $) are the embedding coordinates as introduced earlier in (\ref{e5}) and (\ref{e6}).

We are now going to explore the integrability criteria in the \emph{tensionless} limit \cite{Harmark:2017rpg} of the TNC sigma model (\ref{e12}). This limit is achieved by taking a simultaneous large $ c $ limit of the associated world-sheet d.o.f.. Taking a second scaling limit on the world-sheet fields results in the so called $ U(1) $ Galilean geometry \cite{Harmark:2017rpg} as the target space over which NR strings are propagating\footnote{In case of type IIB (super)strings propagating on $ AdS_5 \times S^5 $ the corresponding dual gauge theory (also known as the Spin-Matrix Theory (SMT)\cite{Harmark:2014mpa}) has been identified as some sort of a decoupling ($ \lambda \rightarrow 0 $) limit of $ \mathcal{N}=4 $  SYM where only states close to the near BPS bound survive \cite{Harmark:2017rpg}. This opens up new possibilities for better understanding of the nonrelativistic holography in the near BPS sector. }. To start with, we scale \cite{Roychowdhury:2019olt}-\cite{Harmark:2019upf} the world-sheet fields as,
\begin{eqnarray}
\lambda = \frac{\mathfrak{g}}{c^2}~;~t= c^{2}\mathfrak{t}~;~\psi =\psi ~;~\theta =\theta ~;~\varphi =\varphi ~;~\zeta =c~ \tilde{\zeta}
\end{eqnarray}
which upon substitution into the original action (\ref{e12}) yields the NR action, $ \tilde{\mathcal{S}}_{NR}=\frac{\sqrt{\mathfrak{g}}}{4 \pi}\int d^{2}\sigma \mathcal{L}_{NR} $ together with the NR Lagrangian,
\begin{eqnarray}
\mathcal{L}_{NR}\approx \frac{\varepsilon^{\alpha \tilde{\alpha}}\varepsilon^{\beta \tilde{\beta}}}{\varepsilon^{\alpha \beta}\partial_{\alpha}\mathfrak{t}\partial_{\beta}\tilde{\zeta}}\partial_{\tilde{\alpha}}\mathfrak{t}\partial_{\tilde{\beta}}\mathfrak{t} ~\mathfrak{h}_{mn}\partial_{\alpha}X^{m}\partial_{\beta}X^{n}+2 \varepsilon^{\alpha \beta}\mathfrak{m}_{\alpha}\partial_{\beta}\tilde{\zeta}+\mathcal{O}(c^{-2}).
\label{e67}
\end{eqnarray}

We choose to work with NR spinning string solitons those are extended along the polar angle ($ \theta $) as well as wrapping and spinning along the azimuthal direction ($ \varphi $) of $ S^2 $,
\begin{eqnarray}
\varphi (\tau , \sigma)=\varpi \tau + \xi (\sigma)~;~\mathfrak{t} = 2 \tau ~;~\tilde{\zeta} = \sigma ~;~ \theta = \theta (\sigma)
\end{eqnarray}
which upon substitution into (\ref{e67}) yields,
\begin{eqnarray}
\mathcal{L}_{NG}=\vartheta'^2_{i} +\xi'^{2}\vartheta^{2}_{i}\delta_{i1}-\Lambda (\vartheta^{2}_{i}-1)+\varpi \vartheta_{2}.
\label{e19}
\end{eqnarray}

The above Lagrangian (\ref{e19}) effectively decribes a one ($ i=1 $) dimensional harmonic oscillator that is constrained to remain on a unit two sphere. This precisely looks like the integrable 1D Rosochatius model \cite{Arutyunov:2016ysi} in the presence of a spin deformation ($ \varpi \vartheta_{2} $) term.  

Given the deformed Lagrangian (\ref{e19}), the first step is to find out the corresponding Hamiltonian dynamics and in particular classify the underlying constraint structure\footnote{The integrability of the reduced 1D model (\ref{e19}) could be anticipated from a naive counting of the integrals of motion associated with the dynamical phase space under consideration. Given a dynamical phase space configuration of dimension $ 2N $ that is subjected to set of (secondary) constraints $ \Psi_{i}(i=1,..,n) $ where, $ n<N $ is said to be \emph{Liouville integrable} if it possess $ \mathcal{I}_{a} (a=N-\frac{n}{2})$ conserved charges those are in \emph{involution}\cite{Arutyunov:2016ysi}. } associated to the deformed 1D Rosochatius model under consideration. In order to do so, we first note down the canonical momenta,
\begin{eqnarray}
\Pi_{1}&=& 2\vartheta'_{1}\\
\Pi_{2}&=&2\vartheta'_{2}\\
\Pi_{\xi}&=&2\xi'\vartheta^{2}_{1}=const.
\end{eqnarray}
which yields the corresponding canonical Hamiltonian density as,
\begin{eqnarray}
\mathfrak{H}_{c}=\Pi^{2}_{i}+\frac{\Pi^{2}_{\xi}}{\vartheta^{2}_{1}}+\tilde{\Lambda} (\vartheta^{2}_{i}-1)+\Delta \mathfrak{h}
\label{e73}
\end{eqnarray}
where we have re-scaled the Hamiltonian (as well as the Lagrange multiplier, $ \Lambda \rightarrow \tilde{\Lambda}=4\Lambda $) by an overall factor of 4 and identify the corresponding deformation piece,
\begin{eqnarray}
\Delta \mathfrak{h}=-4\varpi \vartheta_{2}.
\end{eqnarray} 
Notice that, the deformation $ \Delta \mathfrak{h} $ vanishes as we set the limit, $ \varpi \rightarrow 0 $ and therefore the corresponding Hamiltonian system (\ref{e73}) reduces to the standard Rosochatius like integrable models as discussed in \cite{Arutyunov:2016ysi}. 

Finally, we note down the primary Hamiltonian,
\begin{eqnarray}
\mathfrak{H}_{P}=\mathfrak{H}_{c}+\tilde{\lambda}\phi_{P}
\label{e75}
\end{eqnarray}  
where we identify the associated primary constraint,
\begin{eqnarray}
\phi_{P}= \Pi_{\tilde{\Lambda}}= \frac{\partial \mathcal{L}_{NG}}{\partial \tilde{\Lambda}'} \approx 0.
\label{e76}
\end{eqnarray}

Given the structure (\ref{e75}), we next categorize the class of constraints associated with the dynamical system under consideration. Setting the variation of $\phi_{P}$ equal to zero we find,
\begin{eqnarray}
\Psi_{1}\approx \lbrace \phi_{P} ,\mathfrak{H}_{P}\rbrace_{PB}=\vartheta^{2}_{i}-1 \approx 0.
\label{e77}
\end{eqnarray}

The zero variation of $ \Psi_{1} $ results in the secondary constraint of the following form,
\begin{eqnarray}
\Psi_{2}\approx \lbrace \Psi_{1} ,\mathfrak{H}_{P}\rbrace_{PB}=\Pi_{i}\vartheta_{i} \approx 0.
\label{e78}
\end{eqnarray}

Any further variation of $ \Psi_{2} $ gives back the canonical structure (\ref{e73}) which thereby concludes the chain of constraints in the theory. We are therefore left with two integrals of motion one of which is the canonical momenta ($ \Pi_{\xi} $) conjugate to $ \xi $. The other integral of motion is what we identify below as the generalized conserved charge associated with deformed Rosochatius model constructed above.

Before we get into the integral(s) of motion, it is customary first to introduce Dirac brackets in the context of NR sigma models. We define Dirac bracket between observables in the phase space as,
\begin{eqnarray}
\lbrace \mathcal{A}_{1},\mathcal{A}_{2}\rbrace_{DB}=\lbrace \mathcal{A}_{1},\mathcal{A}_{2}\rbrace_{PB}-\lbrace \mathcal{A}_{1},\Psi_{i}\rbrace (\Gamma^{-1})_{ij}\lbrace \Psi_{j},\mathcal{A}_{2}\rbrace
\end{eqnarray} 
where, $ \Gamma_{ij}=\lbrace \Psi_i , \Psi_j \rbrace_{PB} $.

Below we propose the generalized integral of motion, 
\begin{eqnarray}
\mathfrak{F}_{R}=\frac{1}{2}(\Pi_{i}\vartheta_{j}-\Pi_{j}\vartheta_{i})^2 +\Pi^{2}_{\xi}\frac{\vartheta^{2}_{2}}{\vartheta^{2}_{1}}+ \mathfrak{f}(\Pi_{i},\vartheta_{i})
\end{eqnarray}
subjected to the evaluation of the constraints (\ref{e77}) and (\ref{e78}). We also set,
\begin{eqnarray}
\mathfrak{f}(\Pi_{i},\vartheta_{i}) = -4\varpi \vartheta_{2}
\end{eqnarray}
which thereby yields,
\begin{eqnarray}
\mathfrak{F}_{R}=\mathfrak{H}_{c}-\Pi^{2}_{\xi}.
\end{eqnarray}

Notice that, $  \mathfrak{f}(\Pi_{i},\vartheta_{i})\rightarrow 0 $ in the limit $ \varpi \rightarrow 0 $ and thereby one recovers the original Rosochatius model \cite{Arutyunov:2016ysi}. As a further crosscheck we notice that,
\begin{eqnarray}
\lbrace \mathfrak{F}_{R},\mathfrak{H}_{c}\rbrace_{DB} = \lbrace \mathfrak{H}_{c},\mathfrak{H}_{c}\rbrace_{DB}=0
\end{eqnarray}
together with the fact, $ \lbrace \mathfrak{F}_{R},\Pi_{\xi}\rbrace_{DB} =0 $ which ensures that all the charges are in involution.

To conclude, we notice that nonrelativistic spinning string sigma models on $ R\times S^2 $ could be mapped into (spin)deformed 1D Rosochatius integrable models where one can obtain the corresponding deformed Uhlenbeck integrals \cite{neumann2} and show that those are in involution. There are some further issues that also remains to be explored. It would be really nice to construct the most general class of solutions by introducing so called ellipsoidal coordinates \cite{Arutyunov:2003uj}. The other direction that could possibly be interesting is to construct the nonrelativistic (NR) analogue of the 2D dual version of the rotating string configuration by exchanging the role of world-sheet coordinates \cite{Arutyunov:2003za}. This leads to what is known as NR pulsating string configurations \cite{Roychowdhury:2019olt} with time dependent radial functions. We hope to address some of these issues in the near future.\\

{\bf {Acknowledgements :}}
 The author is indebted to the authorities of IIT Roorkee for their unconditional support towards researches in basic sciences. \\\\ 


\end{document}